# Electrical control of interlayer exciton dynamics in atomically thin heterostructures

Luis A. Jauregui[1], Andrew Y. Joe[1], Kateryna Pistunova[1], Dominik S. Wild[1], Alexander A. High[1,2], You Zhou[1,2], Giovanni Scuri[1], Kristiaan De Greve[1,2], Andrey Sushko[1], Che-Hang Yu[3], Takashi Taniguchi[4], Kenji Watanabe[4], Daniel J. Needleman[3,5,6], Mikhail D. Lukin[1], Hongkun Park[1,2] and Philip Kim[1,3*]

[1] Department of Physics, Harvard University, Cambridge, Massachusetts 02138, USA

[2] Department of Chemistry and Chemical Biology, Harvard University, Cambridge, Massachusetts 02138, USA

[3] John A. Paulson School of Engineering and Applied Sciences, Harvard University, Cambridge, Massachusetts 02138, USA

[4] National Institute for Materials Science, 1-1 Namiki, Tsukuba 305-0044, Japan

[5] Department of Molecular and Cellular Biology, Harvard University, Cambridge, Massachusetts 02138, USA

[6] Faculty of Arts and Sciences Center for Systems Biology, Harvard University, Cambridge, Massachusetts 02138, USA

* To whom correspondence should be addressed: pkim@physics.harvard.edu

**Abstract**

**Excitons in semiconductors, bound pairs of excited electrons and holes, can form the basis for new classes of quantum optoelectronic devices. A van der Waals heterostructure built from atomically thin semiconducting transition metal dichalcogenides (TMDs) enables the formation of excitons from electrons and holes in distinct layers, producing interlayer excitons with large binding energy and a long lifetime. Employing heterostructures of monolayer TMDs, we realize optical and electrical generation of long-lived neutral and charged interlayer excitons. We demonstrate the transport of neutral interlayer excitons across the whole sample that can be controlled by excitation power and gate electrodes. We also realize the drift motion of charged interlayer excitons using Ohmic-contacted devices. The electrical generation and control of excitons provides a new route for realizing quantum manipulation of bosonic composite particles with complete electrical tunability.**

As bosonic composite particles, long-lived excitons can be potentially utilized for the realization of coherent quantum many-body systems (*1, 2*) or as quantum information carriers (*3, 4*). In conventional semiconductors, the exciton lifetime can be increased by constructing double quantum well (DQW) heterostructures, where spatially separated electrons and holes form interlayer excitons (IEs) across the quantum wells (*5-10*). Strongly bound IEs can also be formed in atomically thin DQW. By stacking two single atomic unit cells of TMDs into a van der Waals (vdW) heterostructure, we can engineer a DQW with unprecedented properties. TMD heterostructures, such as MoSe$_2$/WSe$_2$, MoS$_2$/WS$_2$, and MoS$_2$/WSe$_2$ have shown ultrafast charge transfer (*11*), the formation of IEs with a large binding energy of approximately 150 meV (*12*), and diffusion over long distances (*13*). Moreover, the tight binding and small exciton Bohr radius potentially allows for quantum degeneracy of the composite bosons, which may lead to exciton condensation at significantly elevated temperatures compared to, e.g., conventional BECs of cold atoms (*2*).

In our experiments, we fabricate individually electrically contacted optoelectronic devices based on a MoSe$_2$ and WSe$_2$ TMD DQW using hexagonal boron nitride (*h*-BN) encapsulated vdW heterostructures (*14, 15*). Optically transparent electrical gates and Ohmic electrical contacts realized for the individual atomic layers allow us to have complete control of the carrier densities in each TMD of the DQW while maintaining full optical access. The top and bottom insets of Figure 1a show an optical image of a representative device with false-colored top gates and a schematic cross-section, respectively (a detailed device scheme is depicted in the Supplementary Materials (SM), S1). The green and red false-colored gates depict the contact gates for doping the MoSe$_2$ and WSe$_2$ regions, respectively. These contact gates together with the pre-fabricated Pt electrodes provide Ohmic contacts in the WSe$_2$ *p*-channel (*16*).

The presence of the top (optically transparent) and bottom electrical gates, in addition to the separately contacted TMD layers, allows us to control the carrier density in the individual TMD layers as well as the electric field across the TMD heterostructure, $E_{hs}$, using the voltage $V_{tg}$ ($V_{bg}$) applied to the top (bottom) gate. For intrinsic TMD layers (i.e., no free carriers and the chemical potential located within the semiconducting gap), where the heterostructure can be approximated by a thin dielectric slab, $E_{hs}$ is controlled by a gate operation scheme where we apply opposite gate polarity (S2). Figure 1a shows the photoluminescence (PL) spectrum measured at temperature $T$ = 4 K as a function of $E_{hs}$, keeping both TMD layers intrinsic. We observe a linear shift of PL peak energy with $E_{hs}$, suggesting a first order Stark shift due to the static electric dipole moment across the vdW heterostructure. By fitting the linear PL peak shift with the linear Stark shift formula $-edE_{hs}$, where $ed$ is the dipole moment, we estimate $d \approx 0.6$ nm, in good agreement with the expected vdW separation between WSe$_2$ and MoSe$_2$. This analysis strongly suggests that the observed PL peak indeed corresponds to the IE emission of the WSe$_2$/MoSe$_2$ heterostructure with out-of-plane oriented electric dipoles. Similar to a previous study (*12*), we also note that the PL from the intralayer excitons is strongly suppressed compared to that of IEs in the WSe$_2$/MoSe$_2$ heterostructure region, suggesting a fast dissociation of intralayer excitons and an efficient conversion to IEs in this system. We further confirm from the constant normalized reflection $\Delta R/R$ at the intralayer exciton resonances (*17-19*) that our gate operation scheme only varies $E_{hs}$ while keeping the layers intrinsic (Figure 1b).

The IEs in the vdW heterostructure can live longer than intralayer excitons due to the spatial separation of electrons and holes in the heterostructure. Figure 1c shows the IE lifetime τ as a function of $E_{hs}$, measured using time-dependent PL after pulsed laser illumination. The lifetime τ increases as $E_{hs}$ increases, reaching ≈ 600 ns for $E_{hs}$ > 0.1 V/nm, an unprecedentedly long lifetime for excitons in TMDs and an order of magnitude larger than in previous studies (*12, 20*). The observed tunability of τ can be explained by the changes in the electron-hole wavefunction overlap with $E_{hs}$ applied along the IE dipole orientation. Considering the PL intensity modulation shown in Figure 1a together with the measured τ, we also demonstrate that the emission efficiency (η) can be tuned with $E_{hs}$, reaching η ~ 80 % when $E_{hs}$ is aligned against the IE dipole moment, promoting the recombination process (S4 in SM).

The electrostatic condition in our heterostructures is greatly modified if we change the gate operation scheme such that one of the TMD layers is doped electrostatically, introducing free charge carriers. Figure 1e shows the IE PL spectrum following the gating scheme depicted in Figure 1d. In this scheme, as shown in the normalized reflection $\Delta R/R$ at the intralayer exciton resonances (Figure 1f) the carrier density of MoSe$_2$ (WSe$_2$), $n_{2D}$ ($p_{2D}$), changes with positive $V_{tg}$ (negative $V_{bg}$) while the WSe$_2$ (MoSe$_2$) layer remains intrinsic, keeping $E_{hs}$ constant (S5 in SM). We observe several drastic changes in the IE emission spectrum as $V_{tg}$ ($V_{bg}$) increases (decreases) and $n_{2D}$ > 0 ($p_{2D}$ > 0). First, the IE PL peaks exhibit a sudden red shift for *n*- (*p*-) doping of MoSe$_2$ (WSe$_2$). Second, the PL peaks continuously red-shift as doping increases for both *n*- and *p*- sides. Note that in this regime, $E_{hs}$ is fixed as discussed above, thus such shift cannot be explained by the Stark effect. Lastly, the PL intensity diminishes rapidly as doping increases.

Our measurements in the doped regime can be explained by the formation of charged IEs (CIEs) (*21*). As shown in the normalized reflection measured with the same gating scheme, we can identify the *p*/intrinsic, intrinsic/intrinsic, intrinsic/*n* regions by the disappearance of the absorption dips for intralayer excitons in MoSe$_2$ and WSe$_2$ (*22*), which are well aligned with the sudden red shift observed in IE PL (vertical dashed lines). Thus, this jump in energy can be related to the CIE. We note that charged excitons can be referred to trions (*23-26*), three-body bound states, or alternatively, attractive polarons, excitonic states dressed by a polarized fermionic sea (*27, 28*), similar in monolayer TMDs. The value of the observed jump ≈ 10 meV (15 meV) for positive (negative) CIEs is in good agreement with the calculated binding energy of CIEs (*21*). The lifetime of CIEs is ≈ 100 ns near the band edge, decreasing with increasing doping presumably due to additional decay channels enabled by scattering with free carriers (Figure 1g).

We create high densities of IEs by increasing laser power. In particular, for neutral IEs, we observe that the PL emission is shifted to higher energy with increasing power, as shown in Figure 2a, consistent with a mean-field shift due to the repulsive dipole-dipole interaction between oriented IEs. Following the analysis based on a parallel plate capacitance model used for GaAs DQW IEs (*29*), we obtain a lower bound for the IE density of ~ 5x10$^{11}$ cm$^{-2}$ (S6). The high density of long-lived IEs and the large τ observed in our heterostructures can enable transport of IEs across the samples. Figures 2b-d show the spatial map of the IE PL intensity at different laser powers. The PL signal can be detected far away from the diffraction-limited focused laser spot (< 1 μm in diameter). At the highest power, PL can be observed many microns away from the excitation spot, strongly suggesting transport of IEs across the sample. From these maps, we obtain the normalized, radially-averaged PL intensity, $I_{PL}^{norm}(r)$, where $r$ is measured from the center of the diffraction-limited steady state laser spot. Here, the PL intensity is normalized by the value obtained at $r = 0$. As shown in Figure 2e, away from the laser spot, $I_{PL}^{norm}(r)$ decreases rapidly as $r$ increases. We note, however, that at a given $r$, $I_{PL}^{norm}(r)$ increases with $P$ even at a position far away from the laser spot. Similarly, an increase of $I_{PL}^{norm}(r)$ is observed when adjusting $E_{hs}$ to increase τ at fixed $P$. The latter observation is consistent with exciton transport, as longer-lived IEs may travel farther (S7 in SM). The characteristic length $L_D$ for the decaying behavior of $I_{PL}^{norm}(r)$ can be obtained from fitting $e^{-r/L_D}/\sqrt{r/L_D}$ to $I_{PL}^{norm}(r)$ away from the laser excitation spot (dashed lines in Figure 2e), following the 2D diffusion model with a point source (S8 in SM). As shown on the left axis of Figure 2g, we find that $L_D$ increases as $P$ increases, suggesting increased diffusion at high IE density possibly due to exciton-exciton interactions.

We also measured the temporal decay of the PL intensity using a diffraction-limited focused pulsed laser. Figure 2f shows an estimate of the time-dependent exciton population (integrated PL signal along the heterostructure weighted by $r^2$) after a laser pulse with peak power $P$. The time-dependent PL exhibits a faster decay process with characteristic time scale $τ_1$ ~ 10 ns initially, followed by a slower decay process occurring on the time scale $τ_2$ ~ 100 ns, suggesting that there are two different mechanisms for the PL intensity decay. The value of $L_D$ estimated above can be converted to a diffusion constant according to $D = L_D^2/τ$. Two values $D_1$ and $D_2$ are obtained using the short ($τ_1$) and long ($τ_2$) decay times, respectively, as shown in Figure 2h. Since our lifetime measurement uses a pulsed laser where the interaction driven IE diffusion occurs just after the pulse is off when the IE density remains high, $τ_1$ could be more relevant for the IE diffusion than $τ_2$. The value of $L_D$, however, measured in steady state, would be dominated by $τ_2$. Figure 2h shows that $D_2$ is in the range of 0.01 - 0.1 cm$^2$/s, while $D_1$ changes from 0.1 to 1 cm$^2$/s. Both $D_1$ and $D_2$ are increasing with increasing $P$, providing upper and lower bounds for non-linear IE diffusion due to dipolar repulsive interaction (S8 in SM), respectively.

We obtain further evidence for IE diffusion from time-dependent spatial PL maps with a pulsed laser illuminating the center of the sample. We measure the IE PL intensity $I_{PL}(r,t)$ as a function of distance $r$ (referenced to the laser illumination spot) and time $t$ (referenced to the falling edge of the laser pulse). Figures 2i-j show the normalized time-dependent PL $I_{PL}^{norm}(r,t) = I_{PL}(r,t)(r,t)/I_{PL}(r,t)(r = 0,t)$ at different laser peak powers. The time-dependent root-mean-square radius, $r_{rms}(t) = \sqrt{\langle r^2 \rangle}$, computed from $I_{PL}^{norm}(r,t)$ increases rapidly when the laser is on, reaching a steady state within ~ 200 ns. Interestingly, $r_{rms}(t)$ increases again rapidly within 100 ns after the laser is turned off. While the observed dynamics of IEs can be explained by the diffusion of IEs driven by interaction (S8), an alternative scenario involving the diffusion of photoexcited free carriers *(30)* is also possible. Future experimental studies using resonant excitation of IEs can be potentially utilized to distinguish these scenarios.

Unlike the neutral IEs discussed above, CIEs can be manipulated by an in-plane electric field. Figure 3a shows the spatial map of $I_{PL}$ overlaid with the device image when CIEs are optically excited at the center of the sample. Similar to neutral IEs, CIEs generated at the laser illuminated spot can diffuse across the entire sample. We note that both the WSe$_2$ and MoSe$_2$ layers in our device have multiple electrical contacts that can be used to control the lateral electric field. Figure 3b shows the spatial map of the PL intensity normalized as $I_{PL}(V_{ds})/I_{PL}(V_{ds} = 0)$ when applying a bias voltage of $V_{ds}$ = 3 V across the WSe$_2$ layer. We observe that the grounded edge of the sample becomes brighter with increasing $V_{ds}$ (also see Figure 3c for the normalized average emission intensity along the heterostructure channel). This increase in PL at the heterostructure boundary can be explained by drift of CIEs under the applied bias voltage in the channel. The applied bias $V_{ds}$ creates an electric field to tilt the band structure in the direction of the WSe$_2$ channel, driving positive (+) CIEs along the same direction as shown in the schematic diagram in Figure 3d. At the boundary of the heterostructure, however, the +CIE cannot be transported to the WSe$_2$ p-channel because current across the boundary must be preserved. Therefore, the transported +CIEs recombine to turn into a hole in the WSe$_2$ p-channel. We further confirm this picture of CIE transport by changing the doping and $V_{ds}$ polarity (see S9 for details).

Finally, we demonstrate the electrical generation of IEs by free carrier injection using Ohmic contacts in individual TMD layers. Since our heterostructure forms type-II aligned *p*- and *n*- layers, the charge transport across the WSe$_2$ to MoSe$_2$ is expected to show diode-like rectifying behaviors *(31, 32)*. Figure 4a shows interlayer current ($I_{ds}$) versus interlayer bias ($V_{ds}$) curves, whose characteristic can be modulated by $V_{tg}$ and $V_{bg}$. Changing ($V_{tg}$, $V_{bg}$) adjusts the band offset in the type-II heterojunction and their filling. The inset of Figure 4a shows a map of the forward bias current at a fixed bias voltage. One can identify the region in which both WSe$_2$ and MoSe$_2$ layers remain intrinsic, consistent with the absorption spectrum discussed in Figure 1. Interestingly, we find that this *p-n* device generates detectable electroluminescence (EL) at sufficiently high bias. A particularly interesting EL condition occurs when both TMD layers are intrinsic thus allowing electrons and holes to recombine through the formation of IEs. Figures 4b and c show the EL maps of the heterostructure region. The local EL intensity in the heterostructure depends on the local recombination current density, which can be controlled by ($V_{tg}$, $V_{bg}$) (S10). We find that the EL spectrum resembles the PL spectrum in the same ($V_{tg}$, $V_{bg}$) configuration *(31)*. Figure 4d shows EL vs. $E_{hs}$. Similar to the PL shown in Figure 1a, the EL spectrum shifts linearly with

$E_{hs}$, which can be attributed to the IE Stark effect. More direct evidence that the EL process in our heterostructure is mediated through the IE formation by carrier injection is provided by the EL lifetime. Figure 4e shows the EL intensity as a function of time when we pulse $V_{ds}$ at a fixed ($V_{tg}$, $V_{bg}$). We measured the EL at the falling edge of the pulse. Long and short lifetimes of ≈ 150 ns and ≈ 25 ns were obtained for the gate voltages of $V_{tg}$ = 10 V and -10 V with $V_{bg}$ = 16.3 V, corresponding to the neutral IE and charged IE formation regime, respectively (see Figure 4a inset).

The electrical generation of long-lived interlayer excitons provides an electrically driven near-infrared light source with an energy tunability that ranges over several hundreds of meV and spatial control of the emission. Achieving high density IEs without optical excitation could pave a way to realize quantum condensates in solid-state devices. Large valley polarization (*20, 33*) strongly coupled to the spin may also lead to novel optoelectronic devices based on electrically driven CIEs. Such devices could be potentially utilized to explore novel approaches for both classical and quantum information processing by employing their spin degree of freedom.

**ACKNOWLEDGEMENTS**

We thank Eugene Demler for discussions. This work is supported by the DoD Vannevar Bush Faculty Fellowship (N00014-18-1-2877 for PK, N00014-16-1-2825 for HP), AFOSR MURI (FA9550-17-1-0002), NSF and CUA (PHY-1506284 and PHY-1125846 for HP and ML), ARL (W911NF1520067 for HP and ML), the Gordon and Betty Moore Foundation (GBMF4543 for PK) and Samsung Electronics (for PK and HP). This work was performed in part at the Center for Nanoscale Systems (CNS), a member of the National Nanotechnology Coordinated Infrastructure (NNCI), which is supported by the National Science Foundation under NSF award 1541959. CNS is part of Harvard University. D. N. acknowledges support from the NSF grant DBI-0959721. A. S. acknowledges support from the Fannie and John Hertz Foundation and the Paul & Daisy Soros Fellowships for New Americans. K.W. and T.T. acknowledge support from the Elemental Strategy Initiative conducted by the MEXT, Japan and the CREST (JPMJCR15F3), JST.

**Competing interests:** The authors declare no competing interests. **Data and Materials availability:** All data are present in the paper or the supplementary materials and are available in the supplementary data files.

**Figures and captions**

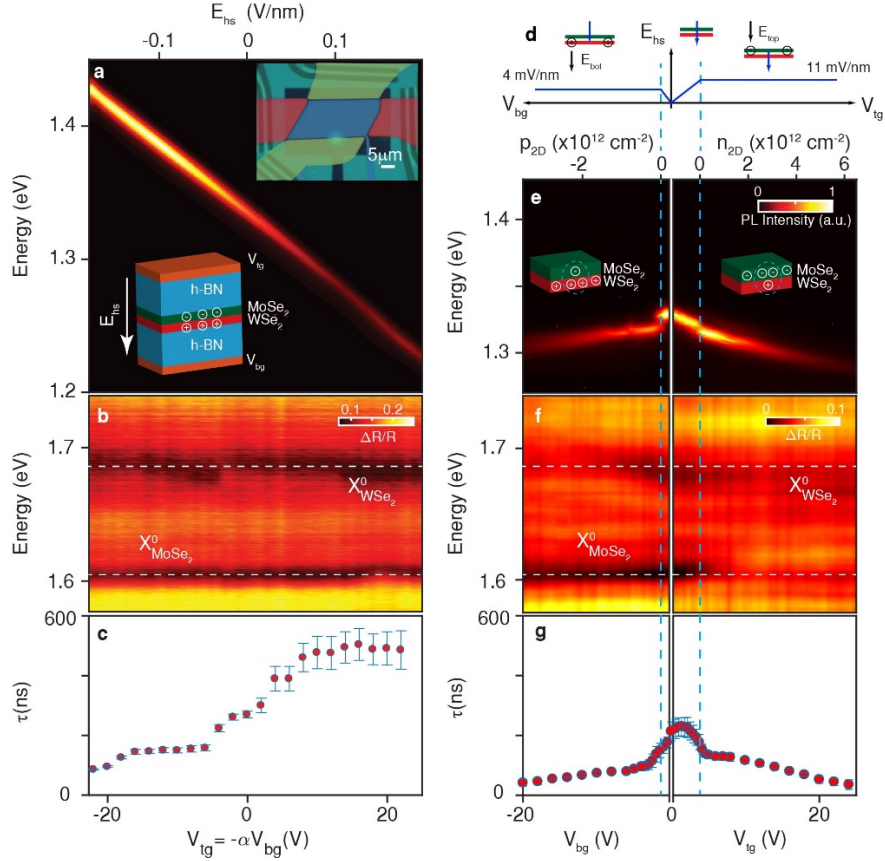

**Figure 1. Electric field & carrier density control of interlayer excitons. (a)** IE PL spectra vs. electric field applied to the heterostructure ($E_{hs} = (V_{tg} - V_{bg})/t_{total} * (\varepsilon_{h-BN}/\varepsilon_{TMD})$). Here the top ($V_{tg}$) and bottom ($V_{bg}$) gate voltages are swept together with a voltage ratio ($\alpha = t_{top}^{h-BN}/t_{bottom}^{h-BN} = 0.614$, $t_{top}^{h-BN} = 70$ nm and $t_{bottom}^{h-BN} = 114$ nm are the top and bottom $h$-BN thicknesses, respectively), $t_{total}$ is the total $h$-BN thickness, and $\varepsilon_{h-BN} = 3.9$ and $\varepsilon_{TMD} = 7.2$ are the $h$-BN and TMD permittivity, respectively. Right inset: optical image of a representative device with the top-gates false colored. Left inset: schematic of the heterostructure cross section, showing electrons (holes) accumulate on the MoSe$_2$ (WSe$_2$) layers, forming IEs. The white arrow represents the positive direction of $E_{hs}$. **(b)** Normalized reflectance vs. $E_{hs}$. **(c)** IE lifetime τ vs. $E_{hs}$. **(d)** Calculated $E_{hs}$ vs. $V_{tg}$ and $V_{bg}$. The field $E_{hs}$ remains constant once a given layer is doped. The top cartoons represent the heterostructure for different applied gate voltages. The fields $E_{hs}$ and $E_{BN}$ (electric field on the top and bottom $h$-BN layers) are depicted as blue and black arrows respectively. **(e)** Single gate dependence ($V_{tg}$ or $V_{bg}$) of the PL shows formation of charged IEs with varying carrier density obtained from the gate operation scheme in (d). Left (right) inset: cartoon of hole (electron)-doped IEs with $V_{bg}$ ($V_{tg}$). **(f)** Normalized reflectance vs. carrier density. The horizontal dashed lines in (b) and (f) represent the neutral excitons for WSe$_2$ ($X_{WSe_2}^0$) and MoSe$_2$ ($X_{MoSe_2}^0$). **(g)** Neutral and charged IE lifetimes τ vs. carrier density. The vertical light blue dashed lines in (d) – (g) mark the intrinsic region.

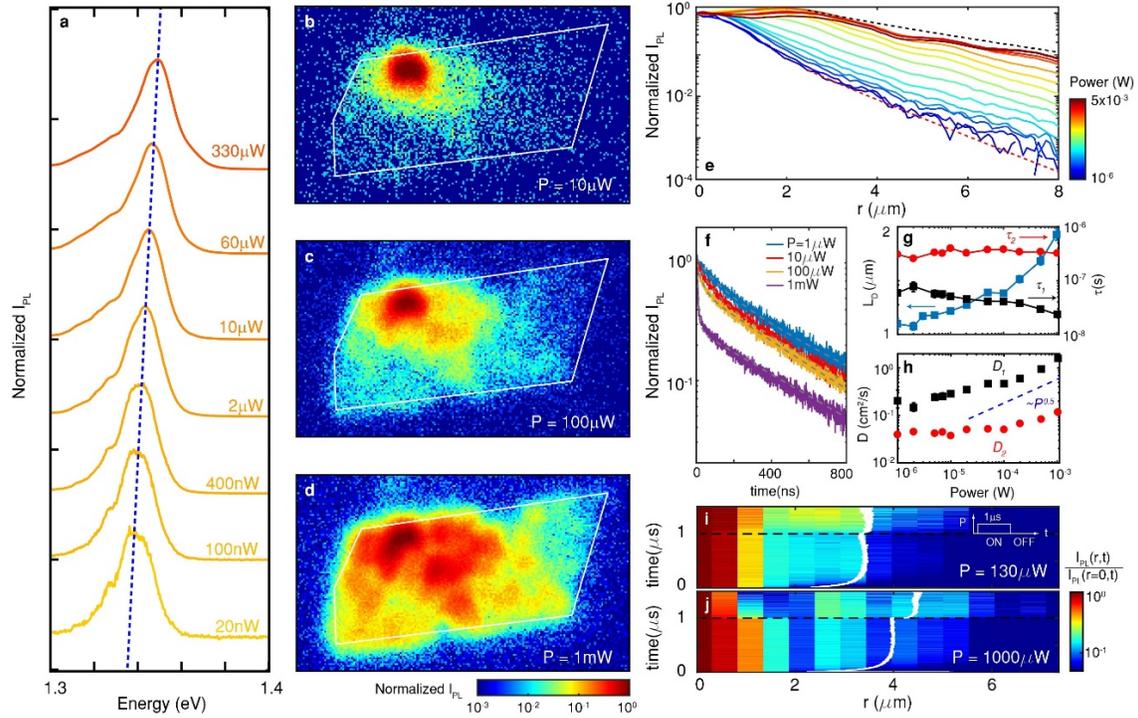

**Figure 2. Spatial control of neutral interlayer excitons. (a)** Power ($P$) dependence of the normalized PL spectra collected from the same spot as the excitation. The blue dashed line corresponds to the PL peak position vs. power. **(b-d)** Spatial dependence of the intensity of the normalized PL for $P$ = 10, 100 and 1000 µW respectively. The continuous wave laser excitation ($\lambda$ = 660 nm) is fixed at the top left of the sample. The white lines depict the heterostructure area. The scale bar corresponds to 5 µm. **(e)** Power dependence of normalized radially averaged $I_{PL}$ (normalized $I_{PL}$) vs. $r$ with the excitation fixed at the center of the sample. The red and black dashed lines represent $e^{-r/L_D}/\sqrt{r/L_D}$ for $L_D$ = 1.1 and 3.2 µm, respectively, where $L_D$ is the diffusion length. **(f)** Time-dependent PL normalized at $t$ = 0 for different ON powers. Dashed gray lines correspond to double exponential fits. **(g)** Left axis: $L_D$ vs. $P$ extracted from Figure 2e. Right axis: lifetime ($\tau$) vs. $P$ with two values of $\tau$ extracted from the double exponential decay fit. **(h)** Diffusion constant ($D = L_D^2/\tau$) vs. $P$ extracted from Figure 2g using the two different values of $\tau$. Dashed line corresponds to $D \sim P^{1/2}$, expected from the non-linear diffusion model (S8). **(i,j)** Normalized $I_{PL}$ vs. time and distance from the laser spot. $I_{PL}(r, t)$ was estimated by averaging over a line cut through the laser spot. We normalize $I_{PL}$ at each time by $I_{PL}(r, t)/I_{PL}(r = 0, t)$. Overlaid white lines represent $\sqrt{\langle r^2 \rangle}$ obtained from the experimental PL map assuming rotational symmetry of the sample. Top right inset: The pulsed laser diode is turned ON at $t$ = 0 µs with powers of 130 µW (i) and 1000 µW (j) is turned OFF at $t$ = 1 µs (also marked with dashed black line).

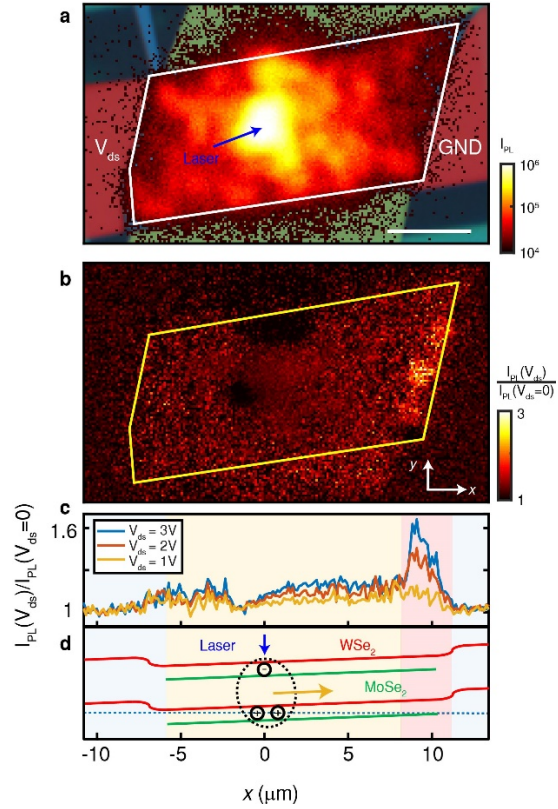

**Figure 3. Spatial control of charged interlayer excitons. (a)** Spatial dependence of $I_{PL}$ with the laser excitation fixed at the center of the heterostructure (laser position labeled as laser). An optical image of the device with false colored top gates that cover the WSe$_2$ and MoSe$_2$ contacts is overlaid. An in-plane electric field is applied by a voltage in one of the WSe$_2$ contacts ($V_{ds}$) while keeping the other contact grounded. **(b)** Spatial dependence of $I_{PL}$ normalized according to $I_{PL}(V_{ds})/I_{PL}(V_{ds} = 0)$ for $V_{ds} = 3$ V. We observe a larger population of charged IEs near the right WSe$_2$ electrode by increasing $V_{ds}$. The yellow arrow in (d) represents the current direction. **(c)** Average of the normalized $I_{PL}$ along the y-axis vs. $x$ (depicted in Figure 3b) for different $V_{ds}$. **(d)** Schematic of the heterostructure bands with applied $V_{ds}$. The red (green) bands correspond to WSe$_2$ (MoSe$_2$). A positive $V_{ds}$ is applied, while the chemical potential (indicated by a blue dotted line) is kept inside the WSe$_2$ valence band to form positively charged IEs. Under positive $V_{ds}$, the CIEs drift towards the grounded contact. The emission mainly occurs near the grounded contact, because the charged exciton cannot move beyond the heterostructure.

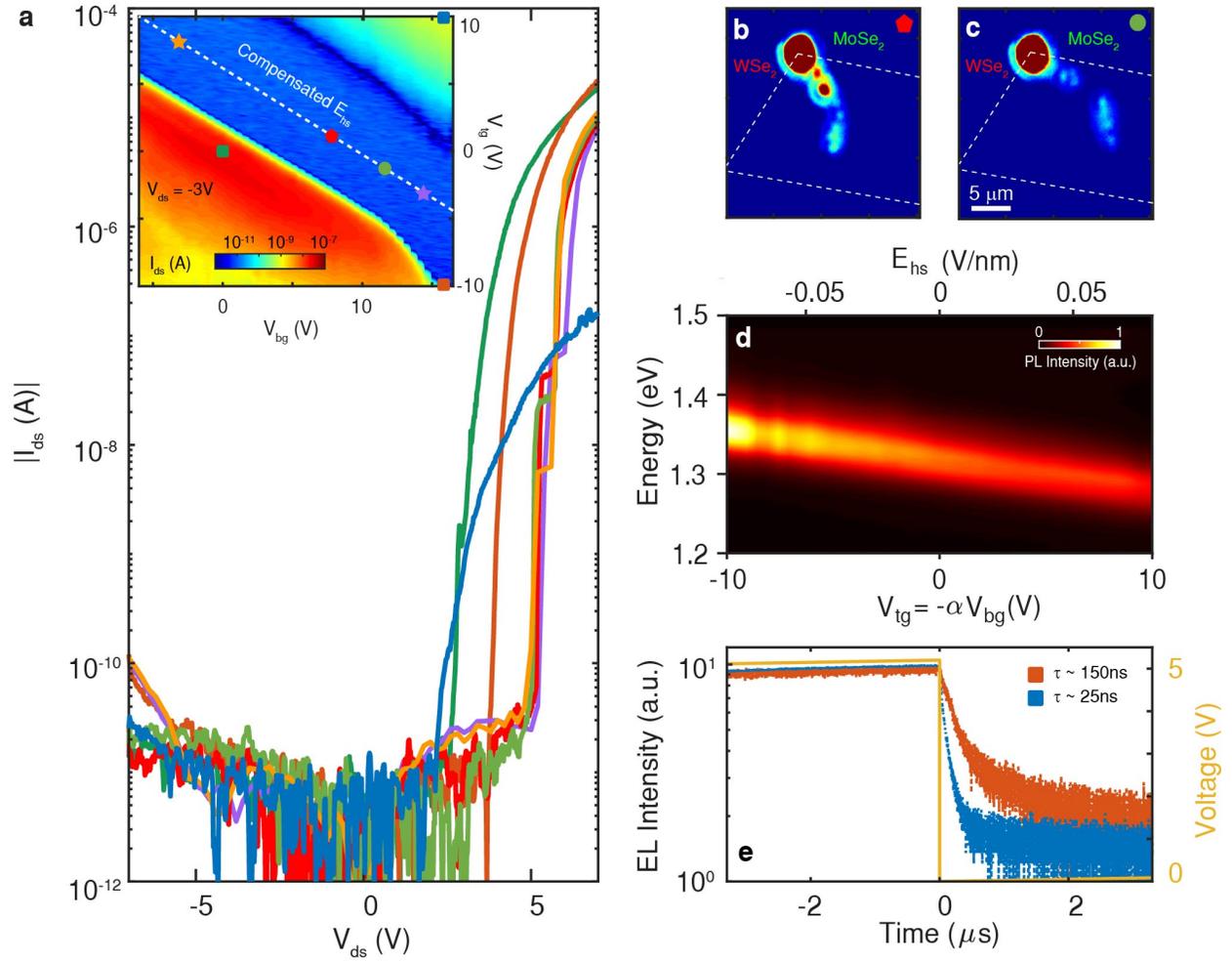

**Figure 4. Electrical generation of interlayer excitons. (a)** I-V curves at various top ($V_{tg}$) and bottom ($V_{bg}$) gate configurations with corresponding indicators in Figure 4a inset. Inset: $I_{ds}$ vs. $V_{tg}$ and $V_{bg}$ (with $V_{ds}$ = -3 V on MoSe$_2$ and grounded WSe$_2$). The white dashed line represents the compensated electric field where $V_{bg} = 10.37V - \alpha V_{tg}$ and $\alpha$ is defined in the Supplementary Information. The markers represent the gate voltages used in (a-c) and (e). **(b)** and **(c)** are spatially dependent electroluminescence (EL) maps for $V_{tg}$ = -1 V (1 V) and $V_{bg}$ = 12 V (8.75 V) at $V_{ds}$ = 7 V. The white dashed lines indicate the heterostructure area. **(d)** Electric field ($E_{hs}$) dependence of the interlayer exciton EL. The electron-hole separation ($d \approx 0.55$ nm) obtained from the slope agrees reasonably with the electron-hole separation extracted from PL. **(e)** Time-dependent EL intensity for two different gate configurations: blue (orange) curve uses $V_{tg}$ = 10 V (-10 V) and $V_{bg}$ = 16.28 V. The yellow line represents the pulsed sawtooth voltage applied to $V_{ds}$. The measured EL lifetime is gate-tunable, as in the PL case, and of comparable magnitude to the PL lifetime.